\documentclass[journal, onecolumn]{IEEEtran}
\ifCLASSINFOpdf
   \usepackage[pdftex]{graphicx}
   \usepackage[skip=2pt, font=footnotesize, labelsep=period]{caption}
   \usepackage{multirow}
   \usepackage{tabularew}
   \usepackage[flushleft]{threeparttable}
   \usepackage{subfig}
   
\else
\fi
\usepackage[cmex10]{amsmath}
\IEEEoverridecommandlockouts

\makeatletter
\renewcommand{\p@subfigure}{\thefigure(}
\makeatletter
\renewcommand{\figurename}{Figure}
\renewcommand\fnum@figure{\textbf{\figurename\nobreakspace\thefigure}}

\usepackage{etoolbox}
\patchcmd{\section}{\centering}{}{}{}

\renewcommand\thetable{\arabic{table}}
\renewcommand\tablename{Table}
\renewcommand\fnum@table{\textbf{\tablename\nobreakspace\thetable}}

\begin{document}
%
\title{MESL: Proposal for a Non-volatile Cascadable \underline{M}agneto-\underline{E}lectric \underline{S}pin \underline{L}ogic}




%


\maketitle
\vspace{-5mm}
\author[{\textbf{Akhilesh Jaiswal$^{1}$$^{,*}$,}}
\author[{\textbf{and Kaushik Roy$^1$}}

\author[{$^1$School of Electrical and Computer Engineering, Purdue University, West Lafayette, IN, 47907, USA}

\author[{$^*$ jaiswal@purdue.edu}
\thispagestyle{plain}
\pagestyle{plain}
\vspace{10mm}

\section*{\large\bf{Abstract}}
\normalsize
In the quest for novel, scalable and energy-efficient computing technologies, many non-charge based logic devices are being explored. Recent advances in multi-ferroic materials have paved the way for electric field induced low energy and fast switching of nano-magnets using the magneto-electric (ME) effect. In this paper, we propose a voltage driven logic-device based on the ME induced switching of nano-magnets.  We further demonstrate that the proposed logic-device, which exhibits decoupled read and write paths, can be used to construct a complete logic family including XNOR, NAND and NOR gates. The proposed logic family shows good scalability with a quadratic dependence of switching energy with respect to the switching voltage. Further, the proposed logic-device has better robustness against the effect of thermal noise as compared to the conventional current driven switching of nano-magnets. A device-to-circuit level coupled simulation framework, including magnetization dynamics and electron transport model, has been developed for analyzing the present proposal. Using our simulation framework, we present energy and delay results for the proposed Magneto-Electric Spin Logic (MESL) gates. 
\\
\setlength{\textfloatsep}{5pt}


%
\IEEEpeerreviewmaketitle


\vspace{-3.5ex}

\section*{\large\bf{Introduction}}

\begin{figure}[h]
\centering
\includegraphics[width=6in]{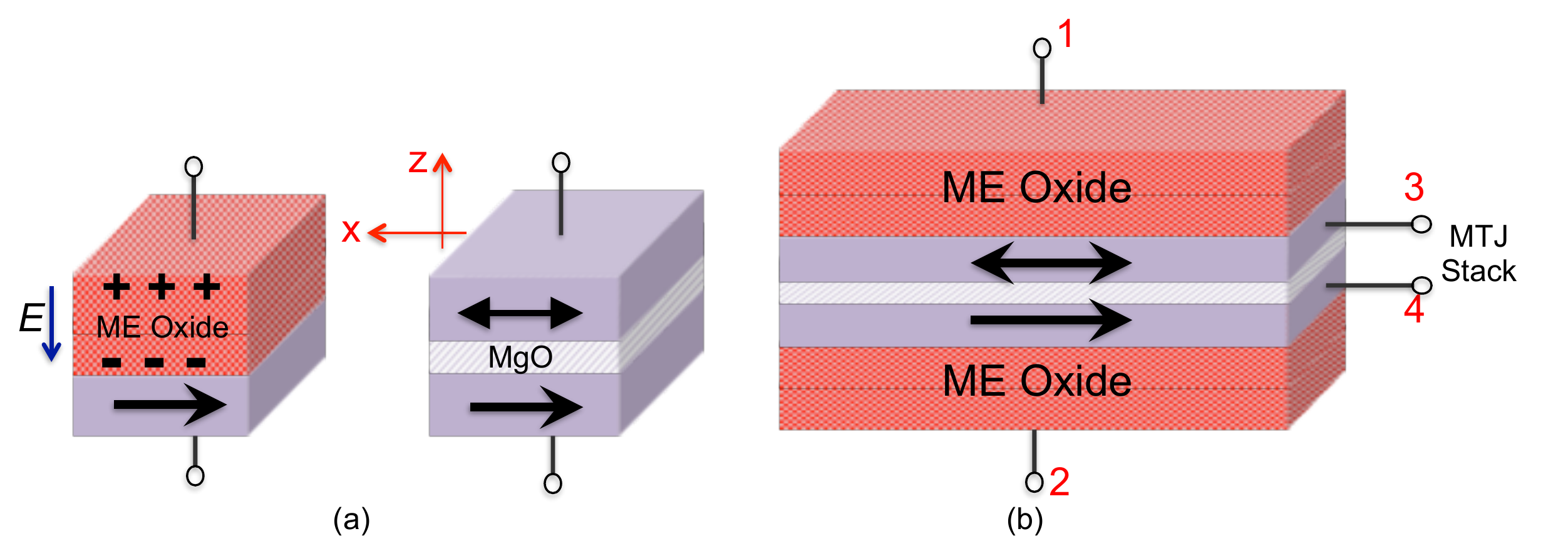}
\caption{\label{fig:epsart} (a) (Left) Figure illustrating the ME switching of a ferro-magnet with applied electric field. A positive voltage on the upper terminal switches the magnet in positive x direction and \textit{vice-versa} (Right) An MTJ stack consisting of an MgO sandwiched between two nano-magnets. The resistance of the MTJ is a function of the voltage and the relative orientation of the magnetization directions. (b) The proposed four terminal logic-device. The upper (lower) nano-magnet can be switched by application of a voltage pulse on terminal 1 (2). The resistance of the MTJ stack can be sensed between terminals 3 and 4. The thickness of the ME oxide and the MgO spacer can be tuned independently to improve the write-efficiency and the sensing margin simultaneously.}
\end{figure}

CMOS technology has been the driving force behind the ever improving computing efficiency for the past few decades \cite{cmos},\cite{cmos2}. However, as the miniaturization of CMOS devices continues, issues like the leakage power consumption, short channel effects, increased variability \textit{etc.} have necessitated exploration of novel devices \cite{leakage}, \cite{tunnel}. Further, with the current emphasis on smart sensors and Internet of Things (IoT), low leakage non-volatile computing devices have became more attractive than ever before. Such beyond-CMOS logic devices are expected to augment/complement the existing CMOS technology \cite{morethanmoore}.

Spin based logic devices are a promising candidate for beyond-CMOS technologies due to 1) non-volatility (ability to retain data in absence of power supply) and hence low leakage power consumption and 2) area-efficiency. As such, many proposals for logical operations using spin devices can be found in the literature. All spin logic (ASL) \cite{ASL_datta}, \cite{ASL_NRL} is one of the widely studied logic families based on non-local spin currents. The dependence of ASL logic on non-local spin currents presents a major drawback due to short \textit{spin-flip} lengths in metallic channels. On the other hand, non-volatile logic based on magnetic tunnel junctions (MTJs) embedded within CMOS logic circuits were explored in \cite{mtjlogic}. In addition, ME based logic devices have been explored in \cite{dc_bridge}, \cite{xor}. The ME logic presented in \cite{dc_bridge} suffers from the requirement of a complex DC (direct current) bridge including three resistors for cascading. In \cite{xor}, an XOR device was proposed, however details of cascading and the complexity of the required cascading circuits is missing. Recently, a spin logic based on magneto-electric switching and the Inverse Rashba Edelstein effect was proposed in \cite{Intel_ME}. In the present paper, we not only demonstrate that our proposed logic-device can function as XNOR, NAND and NOR gate based on the configuration but also show that easy cascadability can be achieved by using minimal number of CMOS devices.

Specifically, we combine two scalable physics, 1) the switching of a ferro-magnet through a multi-ferroic material using the ME effect and 2) the resistance change of an MTJ as a function of the magnetization directions of the constituting ferro-magnets, to propose a non-volatile cascadable Magneto-Electric Spin Logic (MESL). The key highlights of the present work are as follows:

\begin{enumerate}
\item We exploit the inherent coupling of multi-ferroic materials with the underlying magnetization direction of a ferro-magnet to achieve voltage driven low energy switching of nano-magnets.  By stacking two such nano-magnets, in contact with respective ME oxides, we form an MTJ stack. We demonstrate that the resulting device can be used as a logic-element that can be used to implement complex Boolean functions. 
\item Using a coupled magnetization dynamics and electron transport simulation, we show that the proposed logic-device exhibits good scalability, better robustness with respect to the influence of thermal noise and high switching speed as compared to the conventional current driven switching of nano-magnets.
\item Realizations of two input XNOR, NAND and NOR gates, forming a complete logic family, has been demonstrated. Further, we show that the proposed MESL gates can be easily cascaded using a \textit{global-reset} operation and \textit{domino-style} clocking.
\item Typically CMOS logic family requires area expensive storage elements (for example, a flip-flop circuit), in order to retain the output of the logic gates. Such flip-flop circuits become redundant in the proposed MESL gates due to its inherent non-volatility.
\end{enumerate}

\section*{\large\bf{ME Effect and Proposed Logic-Device}}

ME effect is the physics of generating magnetization from an applied electric field \cite{Revival_ME},\cite{revivalme2}. ME devices usually consists of a single phase or composite multi-ferroic material (for example BiFeO$_3$ \cite{Ramesh_ME}, BaTiO$_3$ \cite{Dijken_ME}) in contact with a nano-magnet. Application of an electric field to the multi-ferroic material results in an effective magnetic field experienced by the nano-magnet. If the generated magnetic field is strong enough, the magnetization of the nano-magnet can be reversed. This electric field driven switching of the nano-magnets, shows better energy efficiency and speed, as compared to the classic current induced spin-transfer-torque switching \cite{Slonc}.

In the case of a single phase ME oxide (like BiFeO$_3$), the switching of the nano-magnet due to applied electric field can be explained as follows \cite{Ramesh_ME}. BiFeO$_3$ is a ferro-electric material. Ferro-electricity in BiFeO$_3$ arises due to the shift of Bi$^+$ cations owing to its hybridization with the surrounding oxygen atoms \cite{BFO_story}. The electric polarization of BiFeO$_3$, which is coupled to the (anti) ferromagnetism of the constituent Fe atoms, can be switched by the application of an electric field. Further, the (anti) ferromagnetism of BiFeO$_3$ can be coupled to the ferro-magnetism of an underlying nano-magnet. The magnetization of the nano-magnet can be switched in response to the applied electric field across BiFeO$_3$ \cite{Ramesh_ME}. The various coupling mechanisms that lead to electric-field driven reversal of magnetization direction in the underlying ferromagnet is currently a topic of intense research \cite{Ramesh_ME},\cite{Dijken_ME}.

Nevertheless, the efficiency of the ME effect is usually abstracted by the ME-coefficient denoted as $\alpha_{ME}$ \cite{Ramesh_ME}. $\alpha_{ME}$ is the ratio of magnetic field generated per unit applied electric field. Experimentally, $\alpha_{ME}$ of 1x10$^{-7}$ s m$^{-1}$ has been reported in the literature \cite{Ramesh_ME}. Schematically, an ME switched nano-magnet is shown in Fig. 1(a). When an electric field is applied in the +z direction, the nano-magnet switches to the +x direction due to the ME effect. If the direction of the applied electric field is reversed, the nano-magnet switches to -x direction.

Along side the ME switched magnet in Fig. 1(a), we also show a conventional MTJ stack consisting of an oxide-spacer separating two nano-magnets. A parallel alignment of the magnetization directions in both the nano-magnets results in a low resistance state, while an anti-parallel alignment leads to a high resistance state. The difference in the parallel and anti-parallel resistance is usually indicated by a term called Tunnel Magneto-Resistance (TMR) ratio. In order to maximize the TMR, MgO is usually used as the oxide-spacer in the MTJ stack. The use of MgO as the oxide-spacer can be justified from first principles analysis \cite{why_mgo}, which indicates that the coupling of the Bloch states between the nano-magnet and MgO has an important role in deciding the overall resistance of the MTJ stack.

Thus, we are dealing with two oxides - ME oxide for switching the nano-magnet and MgO for reading the state of the MTJ stack. In Fig. 1(b), we show the proposed device structure. It consists of two nano-magnets in contact with respective ME oxides. Due to their multi-ferroic nature, each of the ME oxides are coupled to the magnetization direction of the underlying nano-magnet. When a positive voltage is applied on the upper or the lower ME oxide, the corresponding nano-magnets switch to +x direction, while for a negative voltage the magnets point in the -x direction. The upper and the lower nano-magnets are separated by an MgO spacer to form an MTJ stack, thus constituting the four-terminal device structure.

The four-terminal nature of the proposed device leads to decoupled read and write paths. Terminals 1 and 2 can be used as the write terminals by applying proper voltage levels to switch the underlying nano-magnets. We assume the ME oxides are thick enough such that the tunneling current flowing through the ME oxides is small enough to be neglected. On the other hand, the state of the device can be read by passing a current (or applying a voltage) between terminals 3 and 4. 
The proposed logic-device thus exhibits 1) low energy consumption due to electric field switching 2) decoupled read/write path such that respective oxides (ME oxide and MgO) can be optimized separately for read and write operations. In the next section, we describe how the proposed logic-device of Fig. 1(b) can be used for constructing  non-volatile XNOR, NAND and NOR gates.

\section*{\large\bf{ME Logic Family and Cascadability}}

\begin{figure}[h]
\centering
\includegraphics[width=5in]{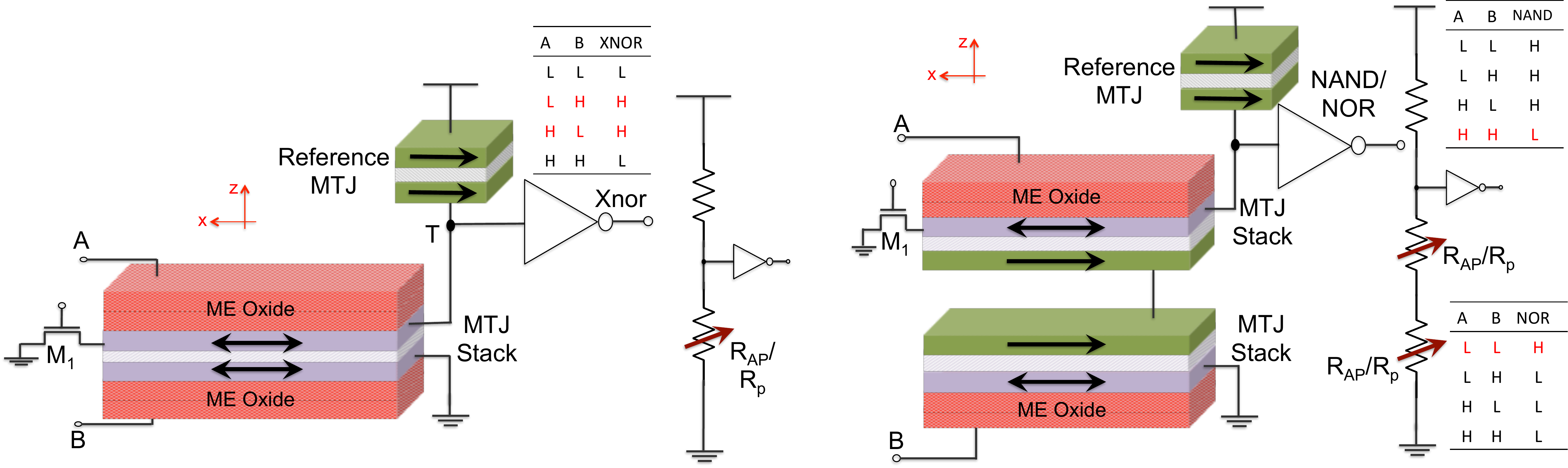}
\caption{\label{fig:epsart} (a)  Proposed ME XNOR gate. Only when both the ferro-magnets point in the same direction, the output of the inverter goes high, thus implementing an XNOR function. Inset shows the truth table for the XNOR function. L represents a digital 0 and H represents a digital 1. (b) Proposed ME NAND/NOR gate. For NAND operation, the inverter is sized such that the output goes low only if both the MTJ stacks are in anti-parallel (high-resistance) state. Whereas, for NOR operation, the sizing of the output inverter is such that it goes high only if both the MTJ stacks are in parallel (low-resistance) state.}
\end{figure}

\begin{figure}[h]
\centering
\includegraphics[width=5in,height=4in]{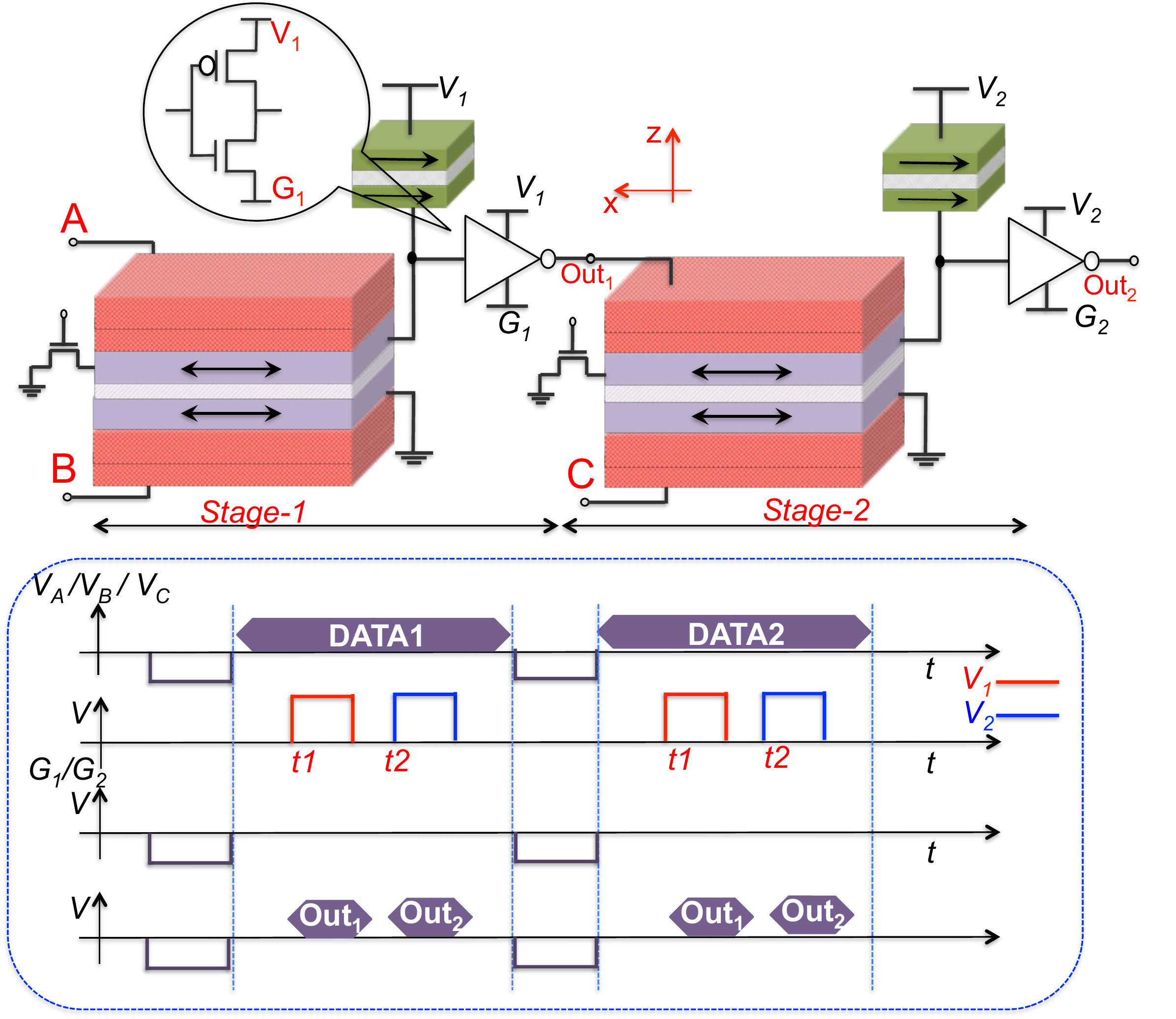}
\caption{\label{fig:epsart} Figure illustrating cascading of two ME XNOR gates. Initially, a reset operation is carried out by applying negative voltage pulses on terminals `A', `B', `C', `G$_1$' and `G$_2$'. On the other hand, when data is applied the two stages are activated in a typical domino-style, one after another. A representative timing diagram illustrates the waveforms on various nodes.}
\end{figure}

The proposed two-input XNOR gate is shown in Fig. 2(a). The inputs to the XNOR gate are terminals `A' and `B'. A positive voltage represents a digital `1' and a negative voltage represents a digital `0'. If, both the inputs are the same, the two nano-magnets will either point in +x direction or in -x direction and the MTJ stack would be in the low resistance (parallel) state. The voltage divider consisting of the reference MTJ and the actual  MTJ stack, will drive the output of the inverter high, if and only if the MTJ stack is in parallel (low resistance) state. Thus, an XNOR function can be implemented using the configuration shown in Fig. 2(a). It is to be noted that while the resistance of the MTJ is being sensed, the voltage divider effect results in a non-zero voltage on the upper ferro-magnet denoted as node `T' in Fig. 2(a) . Since the upper ferro-magnet constitutes one of the plates of the upper ME capacitor, the voltage at node `T' might switch the direction of the upper ferro-magnet. In order to avoid any inadvertent switching of the ferro-magnet, the sensing voltage, the resistance of the reference MTJ and the trip-point of the inverter were selected such that the voltage at node `T' is less than the minimum voltage required to switch the ferro-magnet.

Next, we propose an ME NAND gate as illustrated in Fig. 2(b). The proposed NAND gate is composed of a series connection of two ME logic-devices. Each of the two series connected ME logic-device consists of an ME oxide in contact with a nano-magnet and separated by a fixed magnet using MgO spacer. The two MTJ stacks, shown in Fig. 2(b), switch to the high resistance anti-parallel state, only if the corresponding inputs are high. The output circuit forms a voltage divider as shown on the right hand part of Fig. 2(b). The ratio of the widths of PMOS and NMOS transistors in the output inverter are chosen such that the inverter output goes low if and only if both the MTJ stacks are in the high resistance (anti-parallel) state (or in other words both the inputs `A' and `B' are high). Thus, the circuit in Fig. 2(b) implements a NAND function by proper selection of the transistor widths. Interestingly, the same circuit shown in Fig. 2(b) can also mimic the behavior of a NOR gate. For the NOR gate, the PMOS and NMOS transistors  in the inverter are sized such that, output of the inverter goes high if and only if both the MTJ stacks are in low resistance state (or in other words only if both `A' and `B' are low).

Now that we have all the basic gates for computations, we would present the cascadability of our proposed logic gates. As an example, let us consider, cascading two ME-XNOR gates. As shown in Fig. 3, the output of the first XNOR gate is connected directly to the input of the next XNOR gate. Initially, we do a reset operation by application of a negative voltage pulse so that all the  magnets point in the -x direction. This can be achieved by applying a negative voltage on input terminals `A', `B' and `C', as shown in the timing diagram of Fig. 3. Simultaneously, we pull the `G$_1$' and `G$_2$' terminals of the inverter to negative reset voltages, thus, driving the output of inverters to negative voltages. As such, a global reset can be achieved by simply applying a negative voltage on all the input and the intermediate terminals.

After the reset phase, terminals `G$_1$' and `G$_2$' are kept at zero volts for normal operation.  Data inputs can now be applied on terminals `A', `B' and `C'. Based on the the inputs at `A', `B' and `C' the input magnets would flip if required, making the MTJ stack either high or low resistance. We then apply, voltage pulses on nodes `V$_1$'  and `V$_2$' one after another, in a typical domino style \cite{rabaey}. When a voltage pulse is applied on node `V$_1$', stage 1 (see Fig. 3) evaluates and the node `Out$_1$' goes high or to zero volts. If `Out$_1$' is high, the next stage magnet corresponding to terminal `Out$_1$' switches to +x-direction. If, however, the inputs are such that node `Out$_1$' remains at zero volts, the corresponding next stage magnet does not switch and stays in the desired -x direction. Once the first stage has evaluated, we apply a pulse on terminal `V$_2$', stage-2 evaluates and produces the desired output on `Out$_2$'.

Thus, easy cascadability, is achieved by use of the reset scheme and domino style clocking. Though, clocking is necessary for functioning of the proposed gates, it has been used in almost all non-volatile spin logic to reduce leakage power consumption \cite{ASL_datta}, \cite{Intel_ME}.

\section*{\large\bf{Modeling and Simulation}}
Next, we describe the modeling and simulation framework that was developed to evaluate the proposed ME gates. The device level simulation framework consisted of coupled magnetization dynamics and electron transport model. The required voltage and the time taken for deterministic switching of the nano-magnet due to the ME effect were obtained from stochastic-magnetization dynamics equations including thermal noise. On the other hand, the resistance of the MTJ stack in parallel and anti-parallel state, as a function of the applied voltage was estimated using Non-equilibrium Greens function (NEGF) formalism \cite{negf}.

Under mono-domain approximation, the magnetization dynamics were modeled using the well-know phenomenological equation called the \textit{Landau-Lifshiz-Gilbert} (LLG) equation \cite{llg}. LLG equation can be written as \cite{llg_thesis}  
\begin{equation}
\label{LLG_eqn}
\frac{\partial \widehat{m}}{\partial \tau} = - \widehat{m} \times \vec{H}_{EFF} - \alpha \widehat{m} \times \widehat{m} \times \vec{H}_{EFF}  
\end{equation}
where $\tau$ is $\frac{\vert \gamma \vert }{1 + \alpha^2}t$. In (\ref{LLG_eqn}), $\alpha$ is the Gilbert damping constant, $\gamma$ is the gyromagnetic ratio, $\widehat{m}$ is the unit vector in the direction of the magnetization, $t$ is time and $H_{EFF}$ is the effective magnetic field. $H_{EFF}$ can be written as
\begin{equation}
H_{EFF} = \vec{H}_{demag} + \vec{H}_{interface} + \vec{H}_{thermal} + \vec{H}_{ME}
\end{equation} 
where $\vec{H}_{demag}$ is the demagnetization field due to shape anisotropy. $\vec{H}_{interface}$ is interfacial perpendicular anisotropy, $\vec{H}_{thermal}$ is the stochastic field due to thermal noise and $\vec{H}_{ME}$ is the field due to ME effect.

$\vec{H}_{demag}$ can be written in SI units as \cite{demag}
\begin{equation}
\label{H_demag}
\vec{H}_{demag} = - M_S(\, N_{xx}m_{x}\widehat{x},\,  N_{yy}m_{y}\widehat{y},\,  N_{zz}m_{z}\widehat{z} \,)
\end{equation}
where $m_{x}$, $m_{y}$ and $m_{z}$ are the magnetization moments in x, y and z directions respectively.  $N_{xx}$, $N_{yy}$ and $N_{zz}$ are the demagnetization factors for a rectangular magnet estimated from analytical equations presented in \cite{demag_f}. $M_s$ is the saturation magnetization. The interfacial anisotropy can be represented as \cite{jaiswal_sca}

\begin{figure}[h]
\centering
\includegraphics[width=6in]{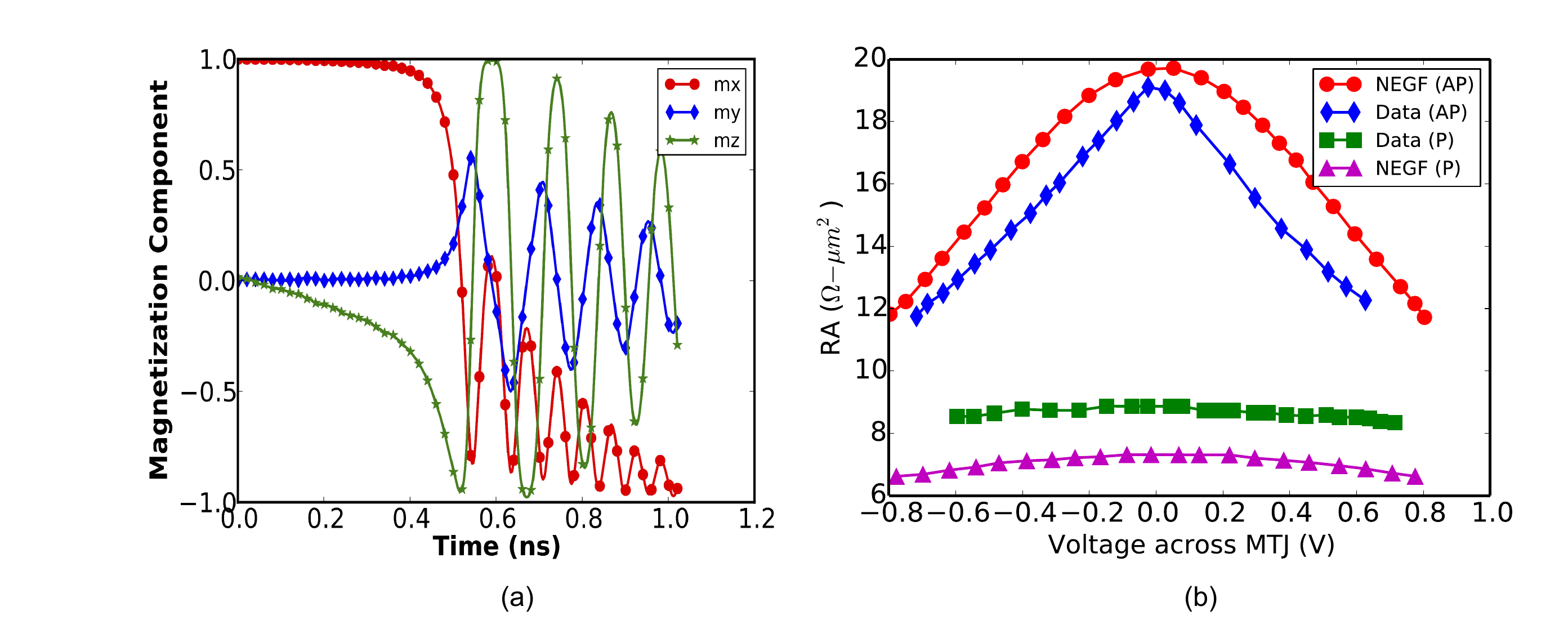}
\caption{\label{fig:epsart} (a) A typical evolution of magnetization components $mx$, $my$, $mz$ on application of a voltage pulse. The magnet is being switched from +x direction to -x direction. (b) The parallel and anti-parallel resistance obtained from our NEGF model \cite{knack} and benchmarked to experimental data from \cite{nnm}. The resistance-voltage characteristics of Fig 4(b), were abstracted into a behavioral model for simulation.}
\end{figure}

\begin{table}[!t]
\renewcommand{\arraystretch}{1.7}
\centering
\caption{Summary of Parameters used for our simulations}
\label{table1}

\begin{tabular}{c c}
\hline \hline
\bfseries Parameters & \bfseries Value\\
\hline
Magnet Length ($L_{mag}$) & $ 45nm \times 2.5 $\\
Magnet Width ($W_{mag}$) & $ 45nm$\\
Magnet Thickness ($t_{FL}$) & $ 2.5nm$\\
ME Oxide Thickness ($t_{ME}$) & $ 5nm$\\
Saturation Magnetization ($M_{S}$) & 1257.3 $KA/m$ \cite{ikeda}\\
Gilbert Damping Factor ($\alpha$) & 0.03 \\
Interface Anisotropy ($K_{i}$) & $1 mJ/m^{2}$ \cite{ikeda}\\
ME Co-efficient ($\alpha_{ME}$) & $0.15/c^* ms^{-1}$ \\
Relative Di-electric constant ($\epsilon_{ME}$) & 500 \cite{Intel_ME}\\
Temperature ($T$) & 300K \\
CMOS Technology & 45nm PTM \cite{PTM} \\

\hline \hline
$*c = Speed \ of \ light.$ 
\end{tabular}
\end{table}

\begin{equation}
\label{H_interface}
\vec{H}_{interface} =  (\, 0\widehat{x},\,  0\widehat{y},\, \frac{2K_{i}}{\mu_{o}M_St_{FL}} m_{z}\widehat{z} \,)  
\end{equation}
where $K_{i}$ is the effective energy density for interface perpendicular anisotropy and $t_{FL}$ is thickness of the free layer. As mentioned earlier, the ME effect can be abstracted through the parameter $\alpha_{ME}$ \cite{Intel_ME} 
\begin{equation}
\vec{{H}}_{ME} = ( \alpha_{ME} (\frac{V_{ME}}{t_{ME}})\widehat{{x}},\,  0\widehat{{y}},\, 0\widehat{{z}} )
\end{equation}
where, $\alpha_{ME}$ is the co-efficient for ME effect, $V_{ME}$ is the voltage applied across the terminals of the ME capacitor and $t_{ME}$ is thickness of the ME oxide, responsible for induction of a magnetic field in response to an applied electric field. $\vec{{H}}_{ME}$ was multiplied with suitable constant for unit conversion.

The thermal field was included by the following stochastic equation \cite{Brown}
\begin{equation}
\label{H_thermal}
\vec{H}_{thermal} = \vec{\zeta}\sqrt{\frac{2\alpha k_BT}{\vert \gamma \vert M_SVol\ dt}}
\end{equation}
where $\vec{\zeta}$ is a vector with components that are zero mean Gaussian random variables with standard deviation of 1. $Vol$ is the volume of the nano-magnet, $T$ is ambient temperature, $dt$ is simulation time step and $k_B$ is Boltzmann's constant. 

Equations (1)-(6) constitute a set of stochastic differential equations. This system of equations was solved numerically by using the Heun's method \cite{Heun}. The solution of the given set of equations, enable us to get the required voltage as well as the switching time for the nano-magnets used in our simulations. The various device dimensions and material parameters used in our simulations are summarized in Table I. A typical evolution of the magnetization components in response to an voltage pulse is shown in Fig. 4(a).

It is to be noted that, multi-ferroics and ME effect is currently an active research area \cite{Revival_ME,revivalme2}. Although many theoretical works have proposed models for describing the origin and the behavior of the ME effect, yet a detailed understanding of the physics of ME effect is still under intense research investigation. Further, experimental demonstration of ME switched ferro-magnets can be found in the literature as in \cite{Ramesh_ME}, yet a global magnetization reversal by ME effect has remained elusive. Due to lack of such experimental results, the ME parameters mentioned in Table I, are not tied to any particular material system or experiment, they are more like predictive parameters that abstract the details of the ME switching into a simple model which can be used in conjunction with the LLG equation to predict the switching time and energy. A more rigorous benchmarking of ME parameters for future logic devices can be found elsewhere as in \cite{nikonov_bench}. Note, the functionality of the present proposal does not depend on the exact values of the ME parameters used for simulation. Therefore, our simple ME model serves the purpose to check the feasibility of our present proposal from device  as well as circuits perspective. 

The magnetization dynamics equations were coupled with the resistance of the MTJ stack, which was modeled using the non-equilibrium Green's function (NEGF) formalism. A detailed description of the NEGF model for MTJs can be found in \cite{knack}. The parallel and anti-parallel resistance obtained from our experimentally benchmarked NEGF equations were then abstracted into a behavioral model. The magnetization dynamics equations along with the resistance of the MTJ stack obtained from the NEGF equations, provided a coupled device model that can be used for characterizing the proposed logic-device.

\section*{\large\bf{Device Characteristics}} 
Any emerging logic technology must exhibit desirable characteristics with respect to scalability, speed of operation \textit{etc}. Using our simulation framework presented in the previous section, we highlight some of the key characteristics of the logic-device proposed in Fig. 1(b) $-$ a) scalability b) stochastic switching dynamics and c) switching speed.

\begin{figure}[h]
\centering
\includegraphics[width=6in]{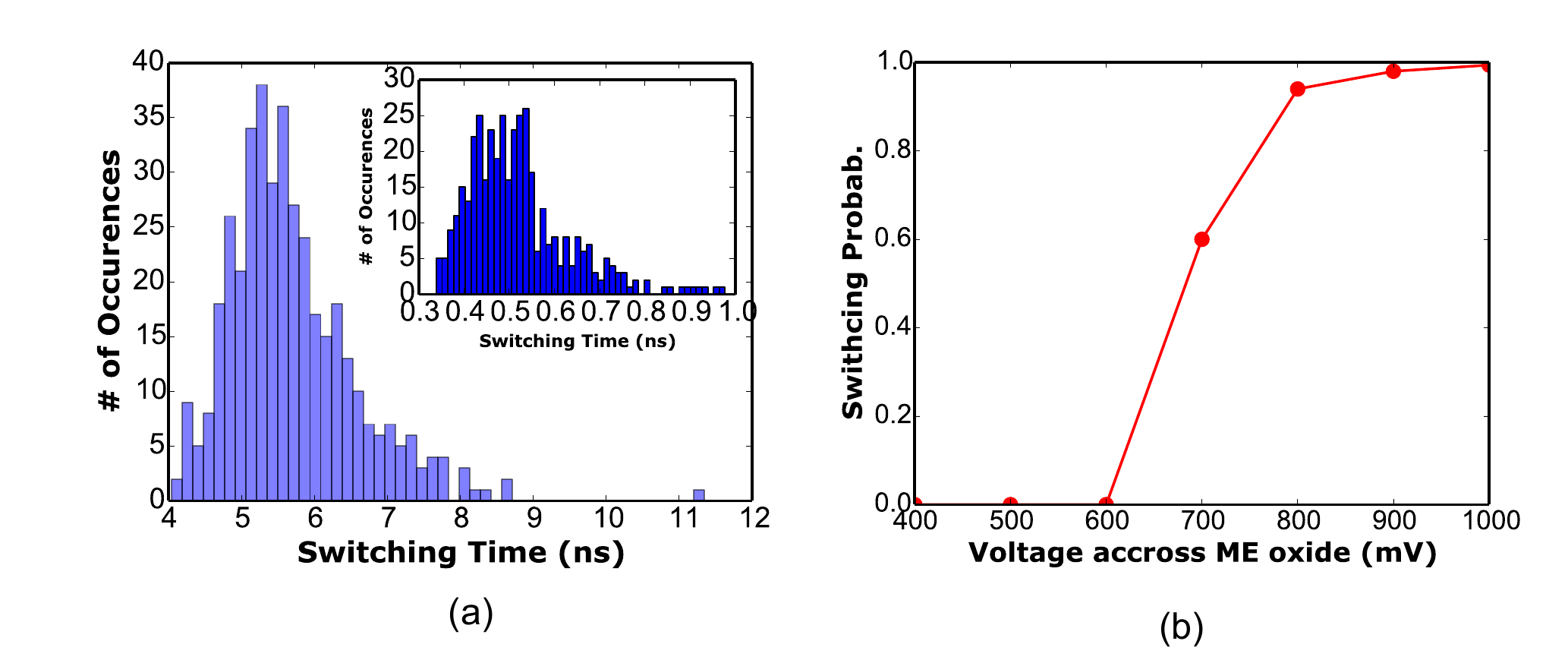}
\caption{\label{fig:epsart} (a)The distribution of switching time for a typical STT based mechanism for constant input current. Inset shows corresponding distribution of switching time for ME based switching for a constant applied electric field. (b) Switching probability as a function of applied voltage across the ME oxide for a switching delay of 500ps. The switching probability has been obtained by running 1,000 LLG simulations with thermal noise. }
\end{figure}

\begin{figure}[h]
\centering
\includegraphics[width=6in]{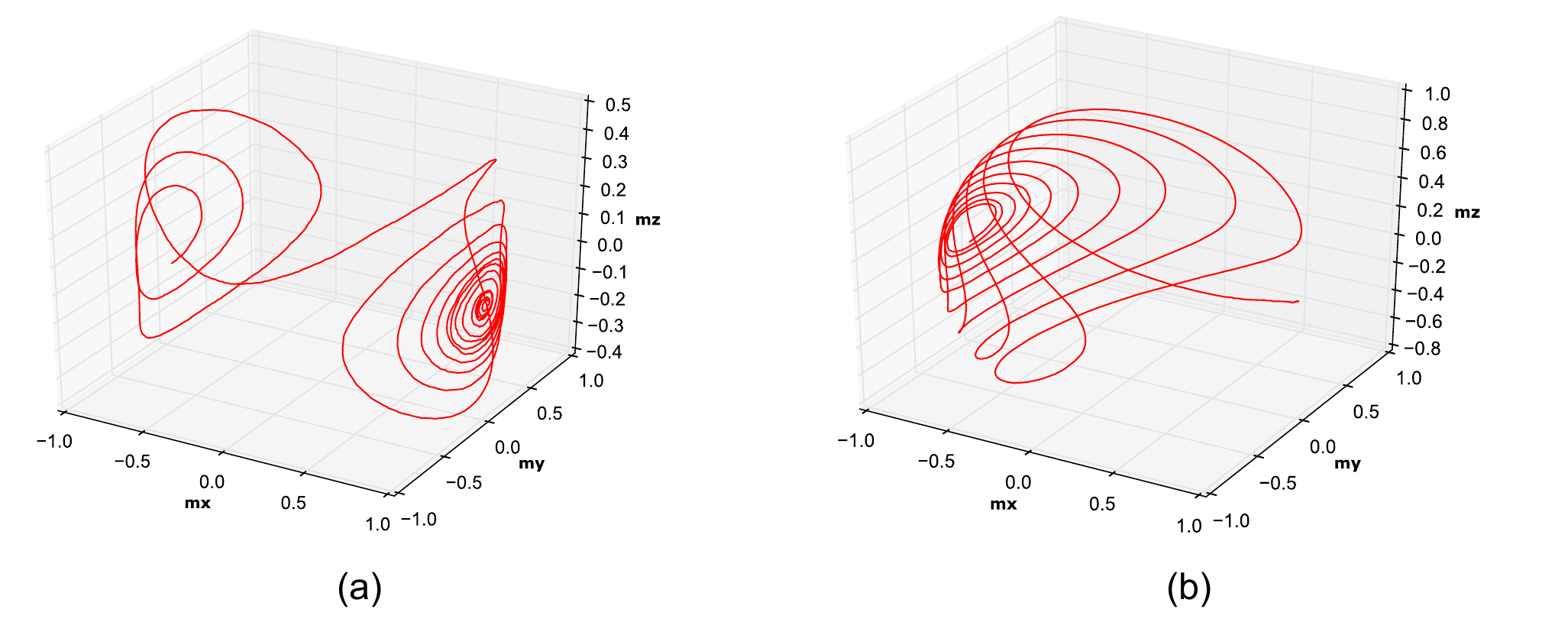}
\caption{\label{fig:epsart} (a)  A typical trajectory followed by the magnetization vector when switched using STT mechanism. The STT mechanism initially acts as an anti-damping torque and subsequently as a damping torque thereby switching the state of the ferro-magnet. (b) A typical trajectory followed by the magnetization vector when switched using the ME mechanism. With application of an external voltage the magnetization tries to orient itself towards the direction of the ME field and finally dampens, resulting in a $180^0$ switching of the ferro-magnet. }
\end{figure}

\subsection{\label{sec:level1} Scalability}
The device in Fig. 1(b) shows good scaling in terms of voltage and energy requirements. This desirable scaling trend can be attributed to the decrease in the ME capacitance  with the decrease in ME oxide area. In fact, the switching energy (proportional to $CV^2$, $C$ being the ME oxide capacitance and $V$ the voltage required to switch the nano-magnet) linearly decreases with the scaling in ME oxide area and has a square law dependence with respect to voltage scaling.

Additionally, the proposed device also requires an MTJ stack for a read-out operation. Recently, many experimental works \cite{mtj17nm},   \cite{mtj20nm} have demonstrated scaling of the MTJ structure to as small as 20nm in diameter. Thus, in terms of areal dimensions of the ME oxide as well as the MTJ stack, the proposed device shows desirable scaling trend.

\subsection{\label{sec:level1} Stochasticity}
The switching dynamics of a nano-magnet is known to be a stochastic process due to the random thermal field as per equation (6). The stochastic switching behavior of the nano-magnets have been exploited for random number generation \cite{rndnum} and in neuromorphic applications \cite{spk}. However, in the present scenario for Boolean logic gates, we need deterministic switching process. Thermal noise has a lesser detrimental effect on ME switched nano-magnets as opposed to the conventional Spin Transfer Torque (STT) switching mechanism. For STT switching, the switching process is initiated by the thermal field. STT effect dominates only when thermal noise has sufficiently disturbed the initial direction of the magnetization vector from its \textit{easy axis} \cite{ther}. On the other hand, ME switching does not require such an initiation of the switching process by the thermal field.

In Fig. 5(a), we have shown the distribution of the switching time for constant current in case of STT switching process and for constant voltage for ME switched nano-magnets. It can be observed that, the spread in distribution of switching time is much lesser for ME switching as compared to the STT switching dynamics. For a given time duration (500ps), the switching probability versus the voltage across the ME oxide is shown in Fig. 5(b). Based on Fig. 5(b), we selected the operating voltage such that the switching probability is $\sim$1.

\subsection{\label{sec:level1} Switching Speed}

ME driven magnetization dynamics can lead to sub-1ns switching speed as compared to the STT mechanism which typically requires 5-10ns of switching time. For STT switching, the STT effect acts as an anti-damping torque initially and as a damping torque subsequently thereby switching the nano-magnet. A typical STT switching curve is shown in Fig. 6(a). On the other hand, ME switching follows a much simpler dynamics, similar to the switching process due an external magnetic field, as shown in Fig. 6(b). For the material parameters shown in Table I, the nano-magnets used in our simulations could switch within 500ps. Thus, as compared to other non-volatile logic devices based on injection of spin current and STT switching mechanism \cite{ASL_NRL}, the proposed logic-device shows faster switching speed. In practice, the total switching time would be the sum of the switching time of the ferro-electric polarization in the ME oxide and the ferro-magnetic switching estimated from our LLG equations.  Theoretically, the ferro-electric switching time can be of the order $\sim$ 70ps \cite{ultra_fast}, which is much less than the ferro-magnetic switching time estimated from our simulations. We have therefore, neglected the ferro-electric switching time in our calculations.

 .

\section*{\large\bf{Results and Discussions}}

The energy associated with a single gate, for example the XNOR gate of Fig. 2(a), can be estimated as follows. The total switching energy would consists of the energy to reset the two nano-magnets, the energy to switch the nano-magnets depending on the incoming data and the energy to turn ON the transistor M1 shown in Fig. 2(a).

\begin{equation}
E_{Swi~Total} = 2C_{ME}V_{Reset}^2  + 2C_{ME}V_{Data}^2  + C_GV_G^2
\end{equation}
where $C_{ME}$ is the capacitance of the ME capacitor, $C_G$ is the gate capacitance of transistor M1, 
$V_{Reset}$ is the reset voltage, $V_{Data}$ is the voltage on terminals `A'/`B' and $V_G$ is the gate voltage for transistor M1. Using our simulation $E_{Swi~Total}$ was estimated to be $5.5fJ$. Similarly, the read-out energy would consists of the energy associated with the voltage divider and the inverter. From our simulations, the read-out energy was estimated to be $30fJ$, assuming a time duration of 500ps.

\section*{\large\bf{Conclusions}}
Non-volatile logic devices are of particular interest given the current emphasis on low power mobile devices, event-driven sensing and Internet of Things. In the present work, we have exploited the ability to switch a ferro-magnet using the ME effect, to propose a non-volatile logic-device. Besides its inherent non-volatility, the present proposal achieves significant benefits in terms of switching energy of the ferro-magnets due to the use of ME effect. Further, the proposed MESL gates can be easily cascaded to implement more complex Boolean functions. From a device perspective, the proposed logic-device shows good scalability, better robustness to thermal fluctuations and high switching speed. We envisage that the proposed MESL gates could be a promising candidate for beyond-CMOS low leakage logic devices. 

\section*{\large\bf{Acknowledgments}}
The work was supported in part by, Center for Spintronic Materials, Interfaces, and Novel Architectures (C-SPIN), a MARCO and DARPA sponsored StarNet center, by the Semiconductor Research Corporation, the National Science Foundation, Intel Corporation and by the DoD Vannevar Bush Fellowship.






 \bibliographystyle{naturemag}
%
%


\end{document}